\documentclass[12pt]{article}

\usepackage{amssymb}
\usepackage{amsmath}
\usepackage{amsbsy}
\usepackage{amsthm}
\usepackage{algorithm,algorithmic}
\usepackage{setspace}
\usepackage{amscd}
\usepackage{amsfonts}
\usepackage{color}
\usepackage{subcaption} 
\usepackage{framed}
\usepackage{mdframed}
\usepackage{adjustbox} 
\usepackage{graphicx}
\usepackage{float}
\usepackage{comment}
\usepackage{fancyhdr}

\newcommand{\LM}[1]{\hbox{\vrule width.2pt \vbox to#1pt{\vfill \hrule width#1pt height.2pt}}}
\newcommand{\LL}{{\mathchoice
{\,\LM7\,}{\,\LM7\,}{\,\LM5\,}{\,\LM{3.35}\,}}}

\graphicspath {
    {./}
}

\begin{document}

\pagestyle{fancyplain}
    \lhead[\fancyplain{}{\sl T.~Kirchdoerfer and M.~Ortiz}]
      {\fancyplain{}{\sl T.~Kirchdoerfer and M.~Ortiz}}
    \rhead[\fancyplain{}{\bfseries\thepage}]
      {\fancyplain{}{\bfseries\thepage}}
    \chead[\fancyplain{}
    {\sl $\qquad\qquad\qquad\qquad$ Data-Driven Computing}]
      {\fancyplain{}
    {\sl $\qquad\qquad\qquad\qquad$ Data-Driven Computing}}
    \cfoot{}

\title{Data-Driven Computing in Dynamics}

\author{
        T.~Kirchdoerfer and M.~Ortiz\\
        Graduate Aeronautical Laboratories \\
        California Institute of Technology \\
        Pasadena, CA 91125, USA}

\maketitle

\begin{abstract}
We formulate extensions to Data Driven Computing for both distance minimizing and entropy maximizing schemes to incorporate time integration. Previous works focused on formulating both types of solvers in the presence of static equilibrium constraints. Here formulations assign data points a variable relevance depending on distance to the solution and on maximum-entropy weighting, with distance minimizing schemes discussed as a special case. The resulting schemes consist of the minimization of a suitably-defined free energy over phase space subject to compatibility and a time-discretized momentum conservation constraint. The present selected numerical tests that establish the convergence properties of both types of Data Driven solvers and solutions.
\end{abstract}


\section{Introduction}

As we transition into an era of data generation and collection, constitutive relations will increasingly come to be characterized by information sets that are {\sl data rich} throughout the regimes of interest. In this new environment, where inference is no longer required, empirical summaries will be necessarily less rich than the data upon which they were based. In these circumstances, modeling finds itself unable to take full advantage of the increasingly large data sets. Ultimately, the assumptions of a model become a restriction on the ability of a calculation to reproduce observed behavior. This lack of predictiveness then leads to unresolvable modeling errors that detract from the quality of the solution. The question then becomes how to move scientific computing beyond the modeling paradigm and have it operate directly on the supplied data sets.

This question is strongly reminiscent of the new field of {\sl Data Science}. In its most general form, Data Science is the extraction of {\sl 'knowledge'} from large volumes of unstructured data \cite{RN176,RN177,RN178,RN417}. It uses analytics, data management, statistics and machine learning to derive mathematical models for subsequent use in decision making. Data Science already provides classification methods capable of processing source data directly into query answers in non-STEM problems. In a similar vein, there has been extensive previous work focusing on the application of Data Science and Analytics to material data sets. The field of Material Informatics (cf., e.~g., \cite{RN211, RN210, RN206, RN205, RN203, RN202, RN200, RN199, RN226, RN227}) uses data searching and sorting techniques to survey large material data sets. It also uses machine-learning regression \cite{RN197, RN212} and other techniques to identify patterns and correlations in the data for purposes of combinatorial materials design and selection. However, what is missing in Material Informatics is an explicit acknowledgement of the field equations of physics and their role in constraining and shaping material behavior. At its best, Material Informatics represents an application of standard sorting and statistical methods to material data sets. While efficient at looking up and sifting through large data sets, it is questionable that any real epistemic knowledge is generated by these methods.

There has also been extensive previous work concerned with the use of empirical data for parameter identification in prespecified material models, or for automating the calibration of the models. For instance, the Error-in-Constitutive-Equations (ECE) method is an inverse method for the identification of material parameters such as the Young's modulus of an elastic material \cite{RN229, RN230, RN231, RN232, RN238, RN241, RN242, RN243, RN244, RN245}. While such approaches are efficient and reliable for their intended application, namely, the identification of material parameters, they differ from Data Driven Computing, as understood here, in that, while material identification schemes aim to determine the parameters of a prespecified material law from experimental data, Data Driven Computing dispenses with material models altogether and uses fundamental material data directly in the formulation of initial-boundary-value problems and attendant calculations thereof.

It has been previously shown \cite{RN48, RN481} that it is indeed possible to reformulate the classical boundary-value problems of mechanics directly in terms of material data alone, without recourse to material modeling, pre-analysis or pre-processing of the material data. In this Data Driven Computing paradigm, the compatibility and conservation laws are recognized as material-independent differential constraints. The Data Driven solution is then defined as the point of the constraint manifold that is closest to the material data set. In this manner, the solution is determined directly by the material data, without recourse to any modeling of the data. This distance-based paradigm has been shown to be well-posed with respect to uniformly convergence of the material-data set \cite{RN48}. The effect of outliers in the material data set can be further mitigated by means of maximum-entropy estimation and information theory \cite{RN481}.

The present work is concerned with the extension of Data Driven computing to dynamics. Distance-minimizing methods described in \cite{RN48} are encompassed as a special case of the applied annealing schedule. Time is discretized using a variational time stepping scheme that is used to generalize the static equilibrium constraints used in previous work. Selected numerical tests are used to demonstrate the convergence properties of both distance minimizing and entropy maximizing data solvers.

The paper is organized as follows. In Section~\ref{P4oupR}, we review max-ent Data Driven solvers and the associated simulated annealing schedules needed for their implementation. In Section~\ref{Wr3uSt} we extend previous Data Driven solvers, concerned with quasistatic problems, to dynamics. In Section~\ref{p6oeWo}, we present numerical tests that assess the convergence properties of max-ent and distance minimizing Data Driven solutions with respect to uniform convergence of the material data set. We also demonstrate the performance of max-ent based Data Driven Computing when the material behavior itself is random, i.~e., defined by a probability density over the phase space. This is followed by a qualitative discussion of performance for these methods, beyond the specifics of convergence. Finally, concluding remarks and opportunities for further development of the Data Driven paradigm are presented in Section~\ref{1iaFRl}.

\section{Review of Data Driven schemes}
\label{P4oupR}

A main task of scientific calculations is to resolve coupled field responses to boundary conditions. Constitutive relationships then define the nature of coupling between the related fields. The language here restricts itself to mechanics, but mechanics is itself a special case of potential field theory through which electrostatics, diffusion and others present a similar need for constitutive definitions. Continuing within mechanics, the relations of interest are the extensive kinematic and kinetic work conjugate fields, e.g. $\varepsilon$ and $\sigma$. Individually these fields must satisfy {\sl material independent} properties with strong constraints. Kinematic fields must satisfy compatibility, while kinetic fields conserve momentum to be consistent with known physical laws. The certainty with which such field constraints can be asserted stands in stark contrast to the {\sl material dependent} constitutive model which typically relates the two fields. Such models must be informed by supplied data, whose summarization into a model is typically performed using ad-hoc empirical fits. These fits, while providing speed and the opportunity for the introduction of intuition and inference, simultaneously introduce a modeling error that influences computational conclusions in way which are hard to characterize.

To move beyond modeling, we now focus on the material data sets upon which such a models are based. This data $E$ exists as a finite point set in phase space $Z$, where an example from small deformation mechanics would express the set as $E=((\varepsilon_i,\sigma_i),i=1,\cdots,N)$. The discrete nature of the set would naturally confound constitutive strategies which rely upon making use of a characterized function form. If compatibility, equilibrium, and boundary conditions are represented by the constraint set $C$, a problem arises in the likely case where the combined constraints cannot be satisfied by couplings defined by the discrete data set, thus $E \cap C$ returns an empty set. What is sought then is a relaxation which continues to satisfy all the members of $C$ while minimizing deviations from $E$ through direct data references.

Initial work on Data Driven computing focused primarily on establishing and demonstrating of a new class of Data Driven solvers \cite{RN48} through the use of a distance minimizing argument. However, such solvers exhibit data-convergence for noisy sets only if the sequence of data sets converges to a graph in the phase space. The need to accommodate a finite band of data obviates the need for a {\sl probabilistic} solution strategy which arbitrates on the relevance and importance of different data points based on proximity. {\sl Cluster analysis} provides a means of incorporating the influence of data neighborhoods to allow data-convergence in the presence of deeper samplings of fixed distributions.

\subsection{Data Clustering}

Data Driven solvers for noisy data have been developed which employ cluster analysis so as to make a new kind of data driven solvers robust to outliers and is well suited to data sources with finite data bands \cite{RN481}. The foundations of cluster analysis have their roots in concepts provided by Information Theory, such as {\sl maximum-entropy} estimation \cite{RN129}. Specifically, we wish to quantify how well a point $z$ in phase space is represented by a point  $z_i$ in a material data set $E=(z_1,\cdots, z_n)$. Equivalently, we wish to quantify the {\sl relevance} of a point $z_i$ in the material data set to a given point $z$ in phase space. We measure the relevance of points $z_i$ in the material data set by means of {\sl weights}  $p_i\in[0,1]$ with the property
\begin{equation}
    \sum_{i=1}^n p_i=1.
\end{equation}
We wish the ranking by relevance of the material data points to be {\sl unbiased}. It is known from Information Theory that the most unbiased distribution of weights is that which maximizes {\sl Shannon's information entropy} \cite{RN132,RN133,RN134}. In addition, we wish to accord points distant from $z$ less weight than nearby points. These competing objectives can be combined by introducing a Pareto weight $\beta\geq0$. The optimal and least-biased distribution is given by the Bolzmann distribution\cite{RN132,RN111}:
\begin{subequations} \label{BoltzDist}
  \begin{align}
      & \label{tR2abr}
      p_i(z,\beta) = \frac{1}{Z(z,\beta)} {\rm e}^{-(\beta/2) d^2 (z,z_i)} ,
      \\ & \label{s5lUsP}
      Z(z,\beta) = \sum_{i=1}^n {\rm e}^{-(\beta/2) d^2 (z,z_i)} .
  \end{align}
\end{subequations}
The corresponding max-ent Data Driven solver now consists of minimizing the free energy
\begin{equation}\label{tpG2dU}
    F(z,\beta) = - \frac{1}{\beta} \log Z(z,\beta) ,
\end{equation}
over the constraint set $C$, i.~e.,
\begin{equation}\label{SoesI5}
    z \in {\rm argmin} \{ F(z',\beta),\ z' \in C \} .
\end{equation}
For finite $\beta$, all points in the material data set influence the solution, but their corresponding weights diminish with distance to the solution. In particular, the addition of an outlier that is marginally closer to the constraint set $C$ than a large cluster of material data points does not significantly alter the solution.

\subsection{Fixed Point Iteration}

Having defined the max-ent Data Driven problem of interest to be the minimization of the free energy $F(z)$ (\ref{tpG2dU}) over the constraint set $C$. The corresponding optimality condition is
\begin{equation}\label{NnS2s3}
    \frac{\partial F}{\partial z}(z,\beta)
    \perp
    C ,
\end{equation}
where $\perp$ denotes orthogonality. Assuming
\begin{equation}\label{4IEbri}
    d(z,z') = | z - z' | ,
\end{equation}
with $| \cdot |$ the standard norm in $\mathbb{R}^n$, we compute
\begin{equation}
    \frac{\partial F}{\partial z}(z,\beta)
    =
    \sum_{i=1}^n
    p_i(z,\beta)
    (z - z_i)
    =
    z
    -
    \sum_{i=1}^n
    p_i(z,\beta)
    z_i .
\end{equation}
Inserting this identity into (\ref{NnS2s3}), we obtain
\begin{equation}
    z
    -
    \sum_{i=1}^n
    p_i(z,\beta)
    z_i
    \perp
    C ,
\end{equation}
which holds if and only if
\begin{equation}\label{AVsY3J}
    z
    =
    P_C\left(
        \sum_{i=1}^n
        p_i(z,\beta)
        z_i
    \right) ,
\end{equation}
where $P_C$ is the closest-point projection to $C$. For instance, if $C = \{ f(z) = 0 \}$ for some constraint function $f(z)$, (\ref{NnS2s3}) may be expressed as
\begin{subequations}\label{uy55LT}
\begin{align}
    &
    \frac{\partial F}{\partial z}(z,\beta)
    =
    \eta \frac{\partial f}{\partial z}(z) ,
    \\ &
    f(z) = 0 ,
\end{align}
\end{subequations}
where $\eta$ is a Lagrange multiplier. We note that eq.~(\ref{AVsY3J}) conveniently defines the following fixed-point iteration,
\begin{equation}\label{uM3Vd6}
    z^{(k+1)}
    =
    P_C\left(
        \sum_{i=1}^n
        p_i(z^{(k)},\beta)
        z_i
    \right) .
\end{equation}
The essential difficulty inherent to problem (\ref{NnS2s3}), or (\ref{uy55LT}), is that, in general, the free energy function $F(\cdot,\beta)$ is strongly non-convex, possessing multiple wells centered at the data points in the material data set. Under these conditions, iterative solvers may fail to converge or may return a local minimizer, instead of the global minimizer of interest. We overcome these difficulties by recourse to {\sl simulated annealing} \cite{RN124}.

\subsection{Simulated Annealing}\label{9iechI}

The general idea of simulated annealing is to evolve the reciprocal temperature jointly with the fixed point iteration according to an appropriate annealing schedule, i.~e., we modify (\ref{uM3Vd6}) to
\begin{equation}
    z^{(k+1)}
    =
    P_C\left(
        \sum_{i=1}^n
        p_i(z^{(k)},\beta^{(k)})
        z_i
    \right) .
\end{equation}
An effective annealing schedule is obtained by selecting $\beta^{(k+1)}$ so as to ensure local contractivity of the fixed-point mapping. An appeal to contractivity \cite{RN481} suggests the schedule
\begin{equation}\label{Vo9Xoa}
    \frac{1}{\beta^{(k+1)}}
    =
    \sum_{i=1}^n
    p_i(z^{(k)},\beta^{(k)})
    | z_i-\bar{z}^{(k)} |^2 ,
\end{equation}
with the initial reciprocal temperature $\beta^{(0)}$ chosen small enough that the mapping $g(\cdot, \beta^{(0)})$ is contractive everywhere. This further leads to an estimate for a convexifying $\beta_0$ with which to initialize the iteration,
\begin{equation} \label{drlE2o}
    \frac{1}{\beta^{(0)}}
    =
    \frac{1}{n}
    \sum_{i=1}^n |z_i|^2
\end{equation}

\section{Application to dynamics}\label{Wr3uSt}

We illustrate the extension of max-ent Data Driven Computing to dynamical problems by means of the simple example of truss structures. Trusses are assemblies of articulated bars that deform in uniaxial tension or compression. Thus, conveniently, in a truss the material behavior of a bar $e$ is characterized by a simple relation between the uniaxial strain $\varepsilon_e$ and uniaxial stress $\sigma_e$ in the bar. We refer to the space of pairs $z_e = (\varepsilon_e, \sigma_e)$ as the {\sl phase space} of bar $e$. We assume that the behavior of the material of each bar $e=1,\dots,m$, where $m$ is the number of bars in the truss, is characterized by---possibly different---local data sets $E_e$ of pairs $z_e$, or {\sl local states}. For instance, each point in the data set may correspond, e.~g., to an experimental measurement. The global data set is then the cartesian product $E = \prod_{e=1}^m E_e$ of all local data sets.

The state $z_k = (z_e)_{e=1}^m$ of the truss at some time $t_k$ is subject to the compatibility and equilibrium constraints
\begin{subequations}
  \begin{align}
      &
      \varepsilon_e = B_e u_k ,
      \\ &
      \sum_{e=1}^m B_e^T w_e \sigma_{e,k} = f_k - M a_k ,
  \end{align}
\end{subequations}
where $u$ and $a$ are the array of nodal displacements and accelerations, $f$ is the array of applied nodal forces, the matrices $(B_e)_{e=1}^m$ encode the geometry and connectivity of the truss members, $w_e$ is the volume of member $e$ and $M$ is the mass matrix. In order to integrate the equations in time we proceed to discretize displacement $u$ and its derivatives $v$ and $a$ in time using the Newmark algorithm
\begin{subequations}
  \begin{align}
    u_{a,k}=u_{a,k-1}+\Delta tv_{a,k-1}+\Delta t^2\left( \left( \frac12-\beta \right)a_{a,k-1}+\beta a_{a,k} \right),\\
    v_{a,k}=v_{a,k-1}+\Delta t\left(1-\gamma)a_{a,k-1}+\gamma a_{a,k} \right),
  \end{align}
\end{subequations}
where $\beta$ and $\gamma$ are the Newmark parameters. In order to reduce these equations to an equivalent static problem, we introduce the Newmark predictors
\begin{subequations}
  \begin{align}
    u^{\text{pred}}_{a,k} &= u_{a,k-1}+\Delta t v_{a,k-1}+\left( \frac12-\beta \right)\Delta t^2 a_{a,k-1},\\
    v^{\text{pred}}_{a,k} &= v_{a,k-1}+(1-\gamma)\Delta ta_{a,k-1},
  \end{align}
  \label{predvar}
\end{subequations}
whereupon the constraints can now be written
\begin{subequations}
  \begin{align}
      \label{Bd2bYR}
      \varepsilon_e & = B_e u_k ,
      \\ \label{j8dCjM}
      \sum_{e=1}^m B_e^T w_e \sigma_{e,k} & = f_k - M\frac{u_{a,k}-u_{a,k}^\text{pred}}{\beta\Delta t^2} ,
  \end{align}
\end{subequations}
with an associated update
\begin{subequations}
  \begin{align}
    &
    a_{a,k} = \frac{u_{a,k}-u_{a,k}^{\text{pred}}}{\beta\Delta t^2},\\ &
    v_{a,k} = v_{a,k}^{\text{pred}}+\gamma\Delta t a_{a,k}.
  \end{align}
  \label{CorrectorUpdate}
\end{subequations}
We may metrize the local phase spaces of each member of the truss by means of Euclidean distances derived from the norms
\begin{equation}\label{fou9Oe}
    | z_e |_e
    =
    \left(
        \mathbb{C} \varepsilon_e^2
        +
        \mathbb{C}^{-1} \sigma_e^2
    \right)^{1/2} ,
\end{equation}
for some positive constant $\mathbb{C}$. We may then metrize the global state of the truss by means of the global norm
\begin{equation}\label{9oakLa}
    | z |
    =
    \Big( \sum_{e=1}^m w_e | z_e |_e^2 \Big)^{1/2}
    =
    \left(
        \sum_{e=1}^m
        w_e
        \Big(
            \mathbb{C} \varepsilon_e^2
            +
            \mathbb{C}^{-1} \sigma_e^2
        \big)
    \right)^{1/2}
\end{equation}
and the associated distance (\ref{4IEbri}). For a truss structure, the point in $C$ closest to a given point $z^*$ in phase space follows from the stationarity condition
\begin{equation}
  \begin{split}
    \delta
    \Bigg\{
        \sum_{e=1}^m
        w_e
        \Bigg(
            \frac{\mathbb{C}}{2} (B_e u_k - \varepsilon_e^*)^2
            & +
            \frac{\mathbb{C}^{-1}}{2} (\sigma_e - \sigma_e^*)^2
        \Bigg)\\
        +
        \Bigg( f & - M \bigg(\frac{u_{k}-u_{k}^{\text{pred}}}{\beta\Delta t^2}\bigg) - \sum_{e=1}^m w_e B_e^T \sigma_e \Bigg)^T \eta
    \Bigg\}
    =
    0 ,
  \end{split}
\end{equation}
where $\eta$ is an array of Lagrange multiplier enforcing the equilibrium constraints. The corresponding Euler-Lagrange equations are
\begin{subequations}
  \begin{align}
    &
    \sum_{e=1}^m w_e B_e^T \mathbb{C} \left(B_{e} u_k- \varepsilon^*_{e,k} \right)
    -M \frac{\eta}{\beta \Delta t^2}=0 \label{eq:uvar},
    \\ &
    \mathbb{C}^{-1}(\sigma_{e,k} - \sigma^*_{e,k}) =
    {B}_{e}\eta,
    \\ &
    \sum_{e=1}^m w_e {B}_{e}^T \sigma_{e,k} = f_k - M \bigg( \frac{ u_{k} - u_{k}^{\text{pred}}}{\beta \Delta t^2}\bigg),
  \end{align}
\end{subequations}
or
\begin{subequations}
  \begin{align}
    &
    \left(\sum_{e=1}^m w_e B_e^T \mathbb{C} B_{e}\right) u_k =
    \sum_{e=1}^m w_e B_e^T \mathbb{C}\ \varepsilon^*_{e,k}
    +M \frac{\eta}{\beta \Delta t^2}
    \\ &
    \left(\sum_{e=1}^m w_e B_e^T \mathbb{C} B_{e}\right) \eta =
    f_{k} - M \frac{u_k-u_k^{\text{pred}}}{\beta \Delta t^2}-\sum_{e=1}^m w_e B^T_e \sigma^*_e
  \end{align}
\end{subequations}
which define two coupled truss equilibrium problems for the linear reference material of modulus $\mathbb{C}$.

\section{Numerical tests}\label{p6oeWo}

\begin{figure}[h]
\begin{subfigure}[]{0.43\textwidth}
    \centering
    \includegraphics[width=\textwidth]{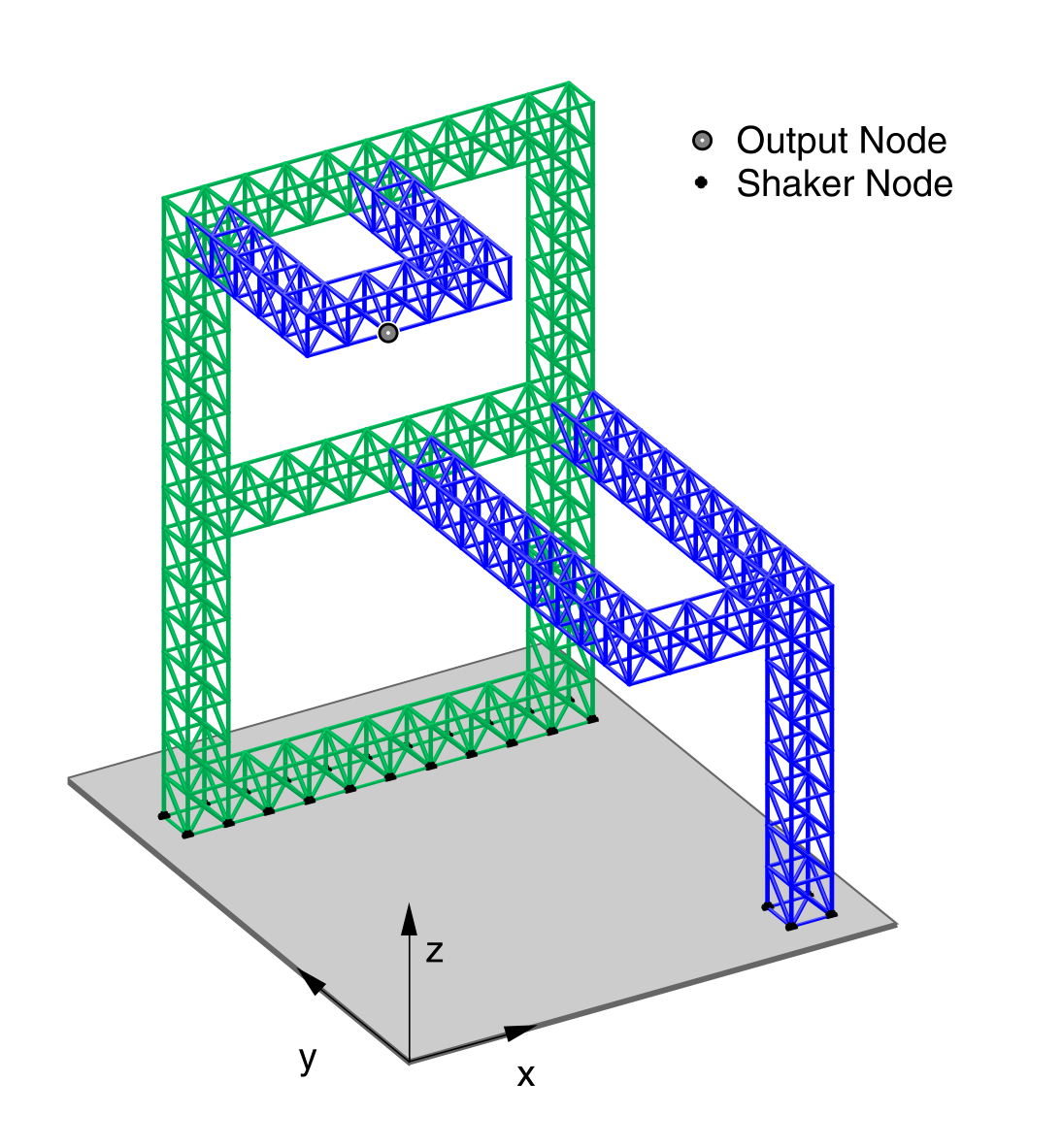}
    \caption{}
\end{subfigure}
\begin{subfigure}[]{0.56\textwidth}
    \centering
    \includegraphics[width=\textwidth]{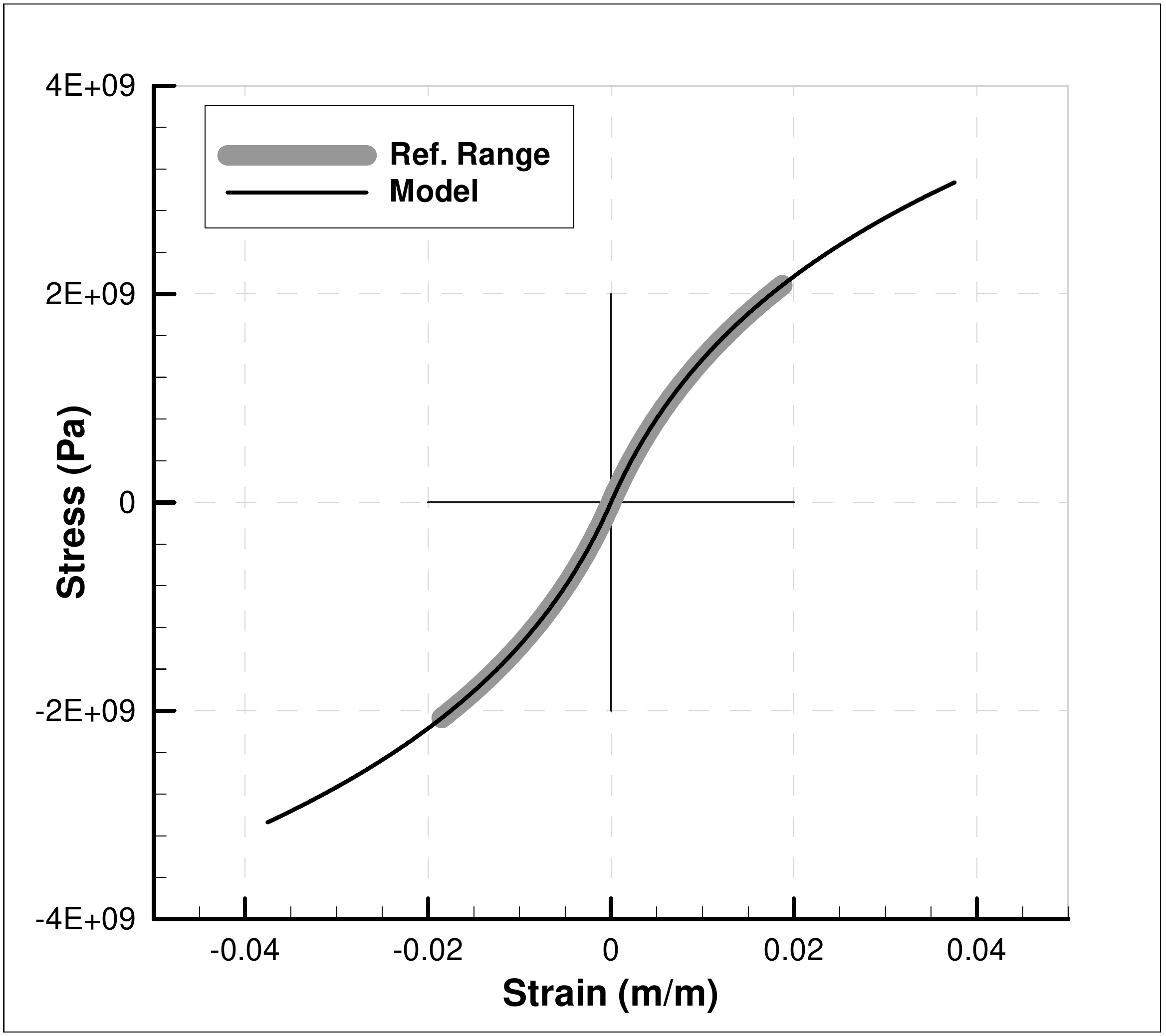}
    \caption{}
\end{subfigure}
    \caption{a) Geometry and boundary conditions of truss test case. b) Base material model with model sampling ranges superimposed.}
    \label{Truss_Model}
\end{figure}

In calculations we consider the specific test case shown in Fig.~\ref{Truss_Model}a. The truss contains 1,246 members and is supported as shown in the figure.  An instantaneous sine excitation of 10 cycles was affected at the base-attached nodes over a time duration resolved with 300 steps. By way of reference, we consider the nonlinear stress-strain relation shown in \ref{Truss_Model}b. A Newton-Raphson based solution with a consistently selected time integrator is readily obtained. The resulting range of states referenced over all the members of the truss in the course of the full time evolution are shown in Fig.~\ref{Truss_Model}a superimposed on the stress-strain curve in order to visualize the coverage of phase space entailed by the reference solution.

To provide a measure of error that summarizes the performance of comparable time integration solutions requires a systematic way of comparing the solution across multiple timesteps without overweighting long-time phase error. Previous work of comparing transient finite element solutions \cite{RN1989} is modified here to create such an analysis error metric,
\begin{equation}
    \text{ERROR}^2=\int_{t_1}^{t_f} \frac1{t^2}\sum_{e=1}^m w_e d^2(z,z^{\text{ref}}) dt,
\end{equation}
where $t_1$ is the time after one step and $t_f$ is the final time. The need for the lower limit of integration to start after the first time step arises from the singularity that would be created by a possible non-zero distance between reference and Data Driven under initial displacement conditions.

\subsection{Annealing Schedule} \label{AnnealingDefinition}

In this work we consider two independent annealing schedules to integrate time, both of which are described by algorithm \ref{alg:Solver}. The first schedule sets $\beta_0\rightarrow\infty$ and all subsequent steps continue to provide full weight to the nearest neighbor in the data set, thus making it consistent with a distance-minimizing scheme. Such schemes have previously been demonstrated only for static mechanics problems \cite{RN48}.

The second schedule is based on maximum entropy. We specifically consider the case in which the behavior of each bar $e$ is characterized by a local material data set $E_e = \{z_{i_e} = (\epsilon_{i_e}, \sigma_{i_e}) \in \mathbb{R}^2,\ i_e=1,\dots,n_e\}$, where $n_e$ is the number of data points in $E_e$, cf.~\cite{RN48}. The global data set is then the Cartesian product
\begin{equation}\label{8riAmI}
    E = E_1 \times \cdots \times E_m .
\end{equation}
A typical point in such a data set is most convenient indexed as $z_{i_1\dots i_m}$, with $i_e=1,\dots,n_e$, $e=1,\dots,m$, instead of using a single index as in Section~\ref{P4oupR}. The partition function (\ref{s5lUsP}) then takes the form
\begin{equation}\label{CoaTr9}
    Z(z,\beta)
    =
    \sum_{i_1=1}^{n_1} \cdots \sum_{i_m=1}^{n_m}
    {\rm e}^{-(\beta/2)\sum_{e=1}^m d^2(z_e,z_{i_e})} ,
\end{equation}
where the local distance is given by (\ref{fou9Oe}). Rearranging terms, (\ref{CoaTr9}) may be rewritten in the form
\begin{equation}
    Z(z,\beta)
    =
    \prod_{e=1}^m
    \left( \sum_{i_e=1}^{n_e} {\rm e}^{-(\beta/2)d^2(z_e,z_{i_e})} \right)
    \equiv
    \prod_{e=1}^m Z_e(z_e,\beta) ,
\end{equation}
and the total free energy evaluates to
\begin{equation}
    F(z,\beta)
    =
    \sum_{e=1}^m
    \left(
        - \frac{1}{\beta} \log Z_e(z_e,\beta)
    \right)
    \equiv
    \sum_{e=1}^m F_e(z_e,\beta) .
\end{equation}
We note that the total free energy is additive with respect to the free energies $F_e(z_e,\beta)$ of the members. Finally, the Bolzmann distribution (\ref{tR2abr}) becomes
\begin{equation}
    p_{i_1,\dots,i_m}(z,\beta)
    =
    \prod_{e=1}^m
    \left(
        \frac{1}{Z_e(z_e,\beta)}
        {\rm e}^{-(\beta/2) d^2 (z_e,z_{i_e})}
    \right)
    \equiv
    \prod_{e=1}^m p_{i_e}(z_e,\beta) .
\end{equation}
In the case of independent local material data sets, eq.~(\ref{8riAmI}), the bound (\ref{Vo9Xoa}) specializes to \cite{RN481}
\begin{equation}\label{fiEcl5}
    \frac{1}{\beta}
    <
    \sum_{e=1}^m
    \left(
        \sum_{i_e=1}^{n_e}
        p_{i_e}(z_e,\beta)
        d^2(\bar{z}_e,z_{i_e})
    \right) .
\end{equation}
Following \cite{RN481}, we exploit this special structure and refine the bound by applying it at the local level, i.~e., by requiring
\begin{equation}\label{pro3SP}
    \frac{1}{\beta_e}
    <
    \sum_{i_e=1}^{n_e}
        p_{i_e}(z_e,\beta_e)
        d^2(\bar{z}_e,z_{i_e}) ,
\end{equation}
$e = 1,\dots,m$, where $1/\beta_e$ represent local temperatures. We can further define an annealing schedule by taking (\ref{pro3SP}) as the basis for local temperature updates
\begin{equation}
    \frac{1}{\beta_e^{(k+1)}}
    =
    \sum_{i_e=1}^{n_e}
        p_{i_e}(z_e,\beta^{(k)})
        d^2(\bar{z}_e^{(k)},z_{i_e}) ,
\end{equation}
with thermal equilibrium subsequently restored by setting the global temperature to
\begin{equation}\label{Q8ecri}
    \frac{1}{\beta^{(k+1)}}
    =
    \sum_{e=1}^m \frac{w_e^{(k+1)}}{\beta_e^{(k+1)}} ,
\end{equation}
with appropriate weights $w_e^{(k+1)}$. In calculations, we specifically choose
\begin{equation}
    w_e^{(k+1)}
    =
    \frac
    {
        {\rm e}^{-\beta_e^{(k)} F_e(\bar{z}_e^{(k)},\beta_e^{(k)})}
    }
    {
        \sum_{e=1}^m
        {\rm e}^{-\beta_e^{(k)} F_e(\bar{z}_e^{(k)},\beta_e^{(k)})}
    }
    =
    \frac
    {
        Z_e(\bar{z}_e^{(k)},\beta_e^{(k)})
    }
    {
        \sum_{e=1}^m Z_e(\bar{z}_e^{(k)},\beta_e^{(k)})
    } .
\end{equation}
Finally, the initial estimate (\ref{drlE2o}) corresponds to setting
\begin{equation}
    p_{i_e}(z_e,\beta) = \frac{1}{n_e} ,
\end{equation}
whereupon (\ref{Vo9Xoa}) becomes
\begin{equation}
    \frac{1}{\beta^{(0)}}
    =
    \sum_{e=1}^m
    \frac{1}{n_e}
    \left(
        \sum_{i_e=1}^{n_e}
        d^2(\bar{z}_e^{(0)},z_{i_e})
    \right) .
\end{equation}
As a further control on the annealing rate we set
\begin{equation} \label{LamdaSchedule}
    \beta^{(k+1)}
    =
    \lambda \tilde{\beta}^{(k+1)} +(1-\lambda)\beta^{(k)} ,
\end{equation}
where $\tilde{\beta}^{(k+1)}$ is the result of applying the update (\ref{Q8ecri}) and $\lambda$ is an adjustable factor.

\begin{algorithm}[H]
\caption{Data-driven solver (one time step)}
\label{alg:Solver}
\begin{algorithmic}
  \REQUIRE Local data sets $E_e=\{z_{i_e},\, i_e=1,\dots,n_e\}$, $B$-matrices $\{B_e,\, e=1,\dots,m \}$, $k=1$, initial displacements and velocities, force vector $f$, parameter $\lambda$.\\ \vspace{6pt}
\STATE 1) compute predictors $u^{\text{pred}}$ and $v^{\text{pred}}$ using eq.~\eqref{predvar}
\STATE 2) Initialize data iteration. Set $j=0$, compute
\begin{equation}
    \bar{z}_e^{(0)}
    =
    z_e^{(0)}
    =
    \frac{1}{n_e} \sum_{i_e=1}^{n_e} z_{i_e},
    \quad
        \frac{1}{\beta^{(0)}}
    =
    \sum_{e=1}^m
    \frac{1}{n_e}
    \left(
        \sum_{i_e=1}^{n_e}
        d^2(\bar{z}_e^{(0)},z_{i_e})
    \right) .
\end{equation}
\STATE 3) Calculate data associations and precalculate for convexity estimate:
\FORALL {$e=1,\dots,m$}
\STATE 3.1) Set $c_{i_e}^{(j)} = \exp\big(-\beta^{(j)} d^2(z_e^{(j)},z_{i_e})\big),\ i_e=1,\dots,n_e$.\\ \vspace{2pt}
\STATE 3.2) Set $Z_e^{(j)}=\sum_{i_e=1}^{n_e} c_{i_e}^{(j)}$. \\ \vspace{2pt}
\STATE 3.3) Set $p_{i_e}^{(j)}=c_{i_e}^{(j)}/Z_e^{(j)},\ i_e=1,\dots,n_e$. \\ \vspace{2pt}
\STATE 3.4) Set $\bar{z}_e^{(j)}=\sum_{i_e=1}^{n_e} p_{i_e}^{(j)} z_{i_e}$. \\ \vspace{2pt}
\STATE 3.5) Set $D_e^{(j)}=\sum_{i_e=1}^{n_e} c_{i_e}^{(j)}d^2(\bar{z}_e^{(j)},z_{i_e})$\\ \vspace{2pt}
\ENDFOR
\STATE \vspace{-10pt}
\STATE 4) Solve: \\ \vspace{-15pt}
\begin{subequations}\label{eq:Truss:alg:EL}
  \begin{align}
    &
    \left(\sum_{e=1}^m w_e B_e^T \mathbb{C} B_{e}\right) u =
    \sum_{e=1}^m w_e B_e^T \mathbb{C}\ \varepsilon^{(j)}_{e}
    +M \frac{\eta}{\beta \Delta t^2}
    \\ &
    \left(\sum_{e=1}^m w_e B_e^T \mathbb{C} B_{e}\right) \eta =
    f - M \frac{u-u^{\text{pred}}}{\beta \Delta t^2}-\sum_{e=1}^m w_e B^T_e \sigma^{(j)}_{e}
  \end{align}
\end{subequations}
for $u$ and $\eta$.\\ \vspace{5pt}
\STATE 5) Progress Schedule:
\STATE \quad 5.1) Set \\ \vspace{-15pt}
\begin{equation}
    \tilde{\beta}^{(j+1)}
    =
    \left(
        \frac{\sum_{e=1}^m D_e^{(j)}}{\sum_{e=1}^m Z_e^{(j)}}
    \right)^{-1} .
\end{equation} \vspace{-8pt}
\STATE \quad 5.2) Set $\beta^{(j+1)} = (1-\lambda)\beta^{(j)} + \lambda\tilde{\beta}^{(j+1)}$. \\ \vspace{5pt}
\STATE 6) Compute local states $z_{e,j}$:
\FORALL {$e=1,\dots,m$}
\STATE \vspace{-10pt}
\begin{equation}\label{eq:Truss:alg:Loc}
   \varepsilon_e^{(j+1)}
    =
    B_e u^{(j+1)} ,
    \qquad
    \sigma_e^{(j+1)}
    =
    \bar{\sigma}_e^{(j+1)} + \mathbb{C} B_e \eta^{(j+1)}
\end{equation}
\vspace{-15pt}
\ENDFOR
\STATE \vspace{-8pt}
\STATE 7) Test for convergence and cycle the time or data iteration:
\IF{$\{z_e^{(j+1)} = z_e^{(j)},\ e=1,\dots,m\}$}
\STATE {\bf exit}
\ELSE
\STATE $j \leftarrow j+1$,
\STATE {\bf goto} (3).
\ENDIF
\end{algorithmic}
\pagebreak
\end{algorithm}

\subsection{Uniform convergence of a noisy data set towards a classical material model}

\begin{figure}[h]
\begin{subfigure}[]{0.49\textwidth}
    \centering
    \includegraphics[width=\textwidth]{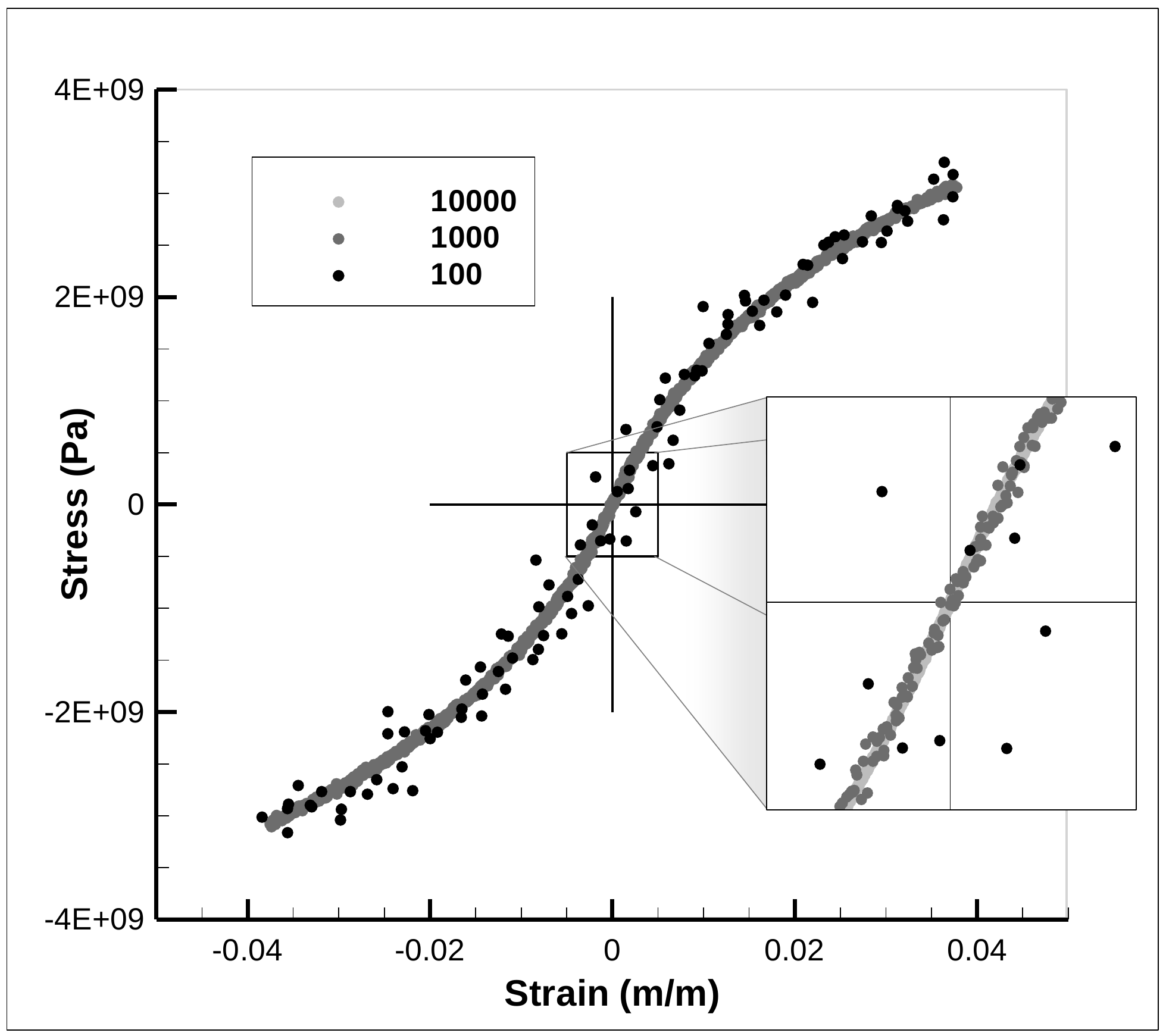}
    \caption{}
\end{subfigure}
\begin{subfigure}[]{0.49\textwidth}
    \centering
    \includegraphics[width=\textwidth]{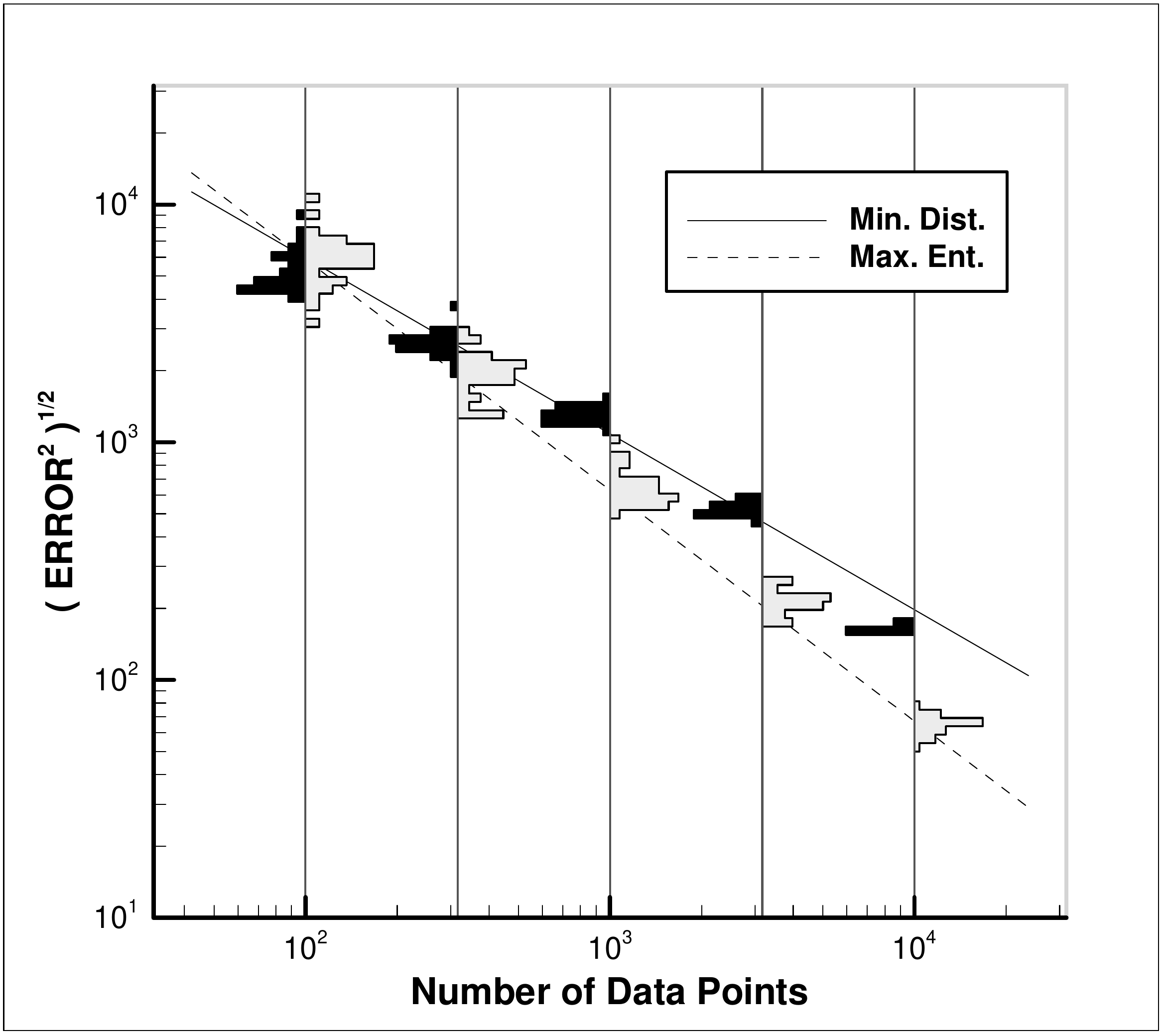}
    \caption{}
\end{subfigure}
    \caption{Truss test case. a) Random data sets generated according to capped normal distribution centered on the material curve of Fig.~\ref{Truss_Model}b with standard deviation in inverse proportion to the square root of the data set size. b) Convergence with respect to data set size of error histograms generated from 30 material set samples.}
    \label{Uniform_Conv}
\end{figure}

Next we consider data sets that, while uniformly convergent to a material curve in phase space, include noise in inverse proportion to the square root of the data set size. To construct a data set consistent with this aim, points are first generated directly from the material curve so that the metric distance between the points is constant. This first sample then has noise added independently pointwise according to a capped normal distribution in both the strain and stress axes with zero mean and standard deviation in inverse proportion to the square root of the data set size. The resulting data sets converge uniformly to the limiting material curve with increasing number of data points. Fig.~\ref{Uniform_Conv}a illustrates the data sets thus generated when the limiting model is as shown in Fig.~\ref{Truss_Model}b.

A convergence plot of error {\sl vs}.~data set size in shown in Fig.~\ref{Uniform_Conv}b, with error metric comparing the max-ent Data Driven solution and the classical solution. For every data set size, the plot depicts histograms of error compiled from 30 randomly generated data set samples. We see that, given the capped structure of the data sets under consideration, distance-minimizing Data Driven solutions converge to the limiting classical solution as $N^{-1/2}$, with $N$ the size of the data set \cite{RN48}. An analysis of Fig.~\ref{Uniform_Conv}b suggests that distance-minimizing schemes convergence rate of $N^{-1/2}$ is consistent with methods employed for static analysis \cite{RN48}. Similarly, max-ent Data Driven solutions converge with a linear rate with respect to the data set size seen in comparable work performed in static analysis \cite{RN481}.

\subsection{Random data sets with fixed distribution about a classical material model}

A different convergence scenario arises in connection with random material behavior described by a {\sl fixed} probability measure $\mu$ in phase space. Specifically, given a set $E$ in phase space, $\mu(E)$ is the probability that a fair test return a state $z \in E$. By virtue of the randomness of the material behavior, the solution becomes itself a random variable. We recall that the constraint set $C$ is the set of states $z$ in phase space that are compatible and in equilibrium. When the material behavior is random and is characterized by a probability measure $\mu$ in phase space, the solution must be understood in probabilistic terms and may be identified with the conditional probability $\mu \LL C$ of $\mu$ conditioned to $C$. The corresponding question of convergence then concerns whether the distribution of Data Driven solutions obtained by sampling $\mu$ by means of data sets of increasing size converges in probability to $\mu \LL C$.

\begin{figure}
\begin{subfigure}[]{0.47\textwidth}
  \centering
  \includegraphics[width=\textwidth]{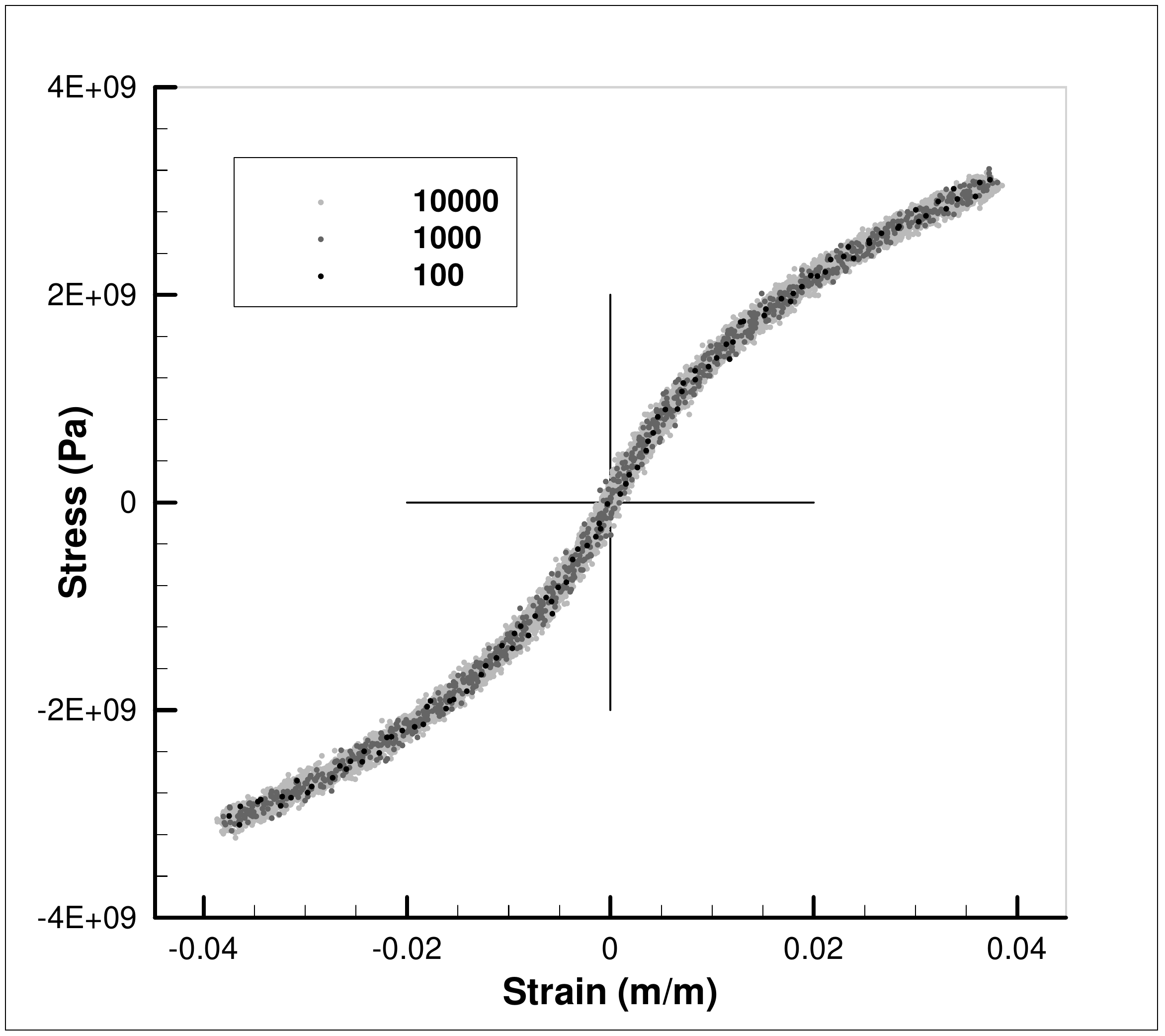}
    \caption{}
\end{subfigure}
\begin{subfigure}[]{0.47\textwidth}
  \centering
  \includegraphics[width=\textwidth]{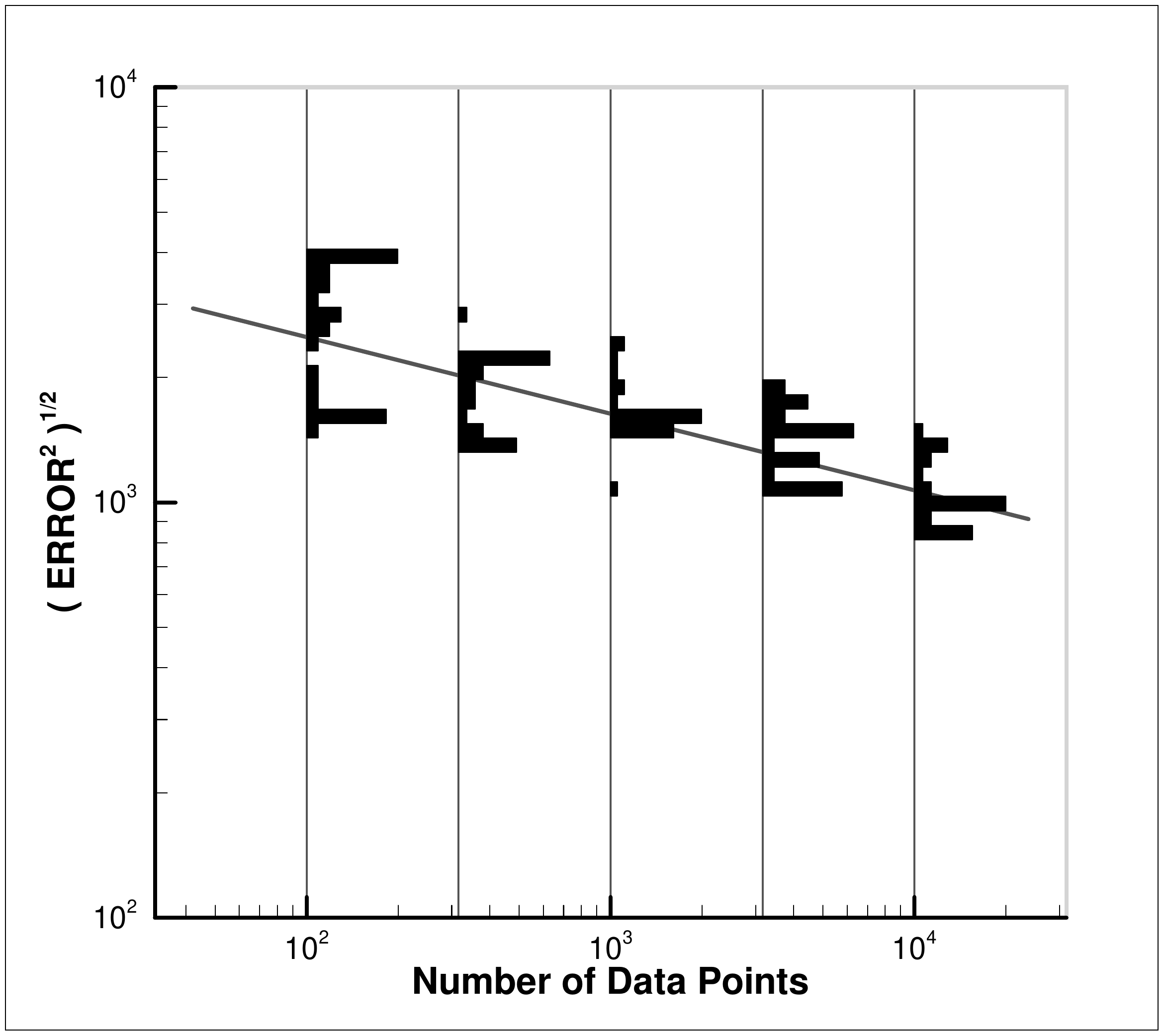}
    \caption{}
\end{subfigure}
  \caption{Truss test case. a) Random data sets generated according to normal distribution centered on the material curve of Fig.~\ref{Truss_Model}b with constant standard deviation independent of the data set size. b) Convergence with respect to data set size of error histograms generated from 30 material set samples.}
  \label{NonUniform_Anneal}
\end{figure}

While a rigorous treatment of convergence in probability is beyond the scope of this paper, we may nevertheless derive useful insights from numerical tests. We specifically assume that $\mu$ is the cartesian product of member-wise measures $\mu_e$ characterizing the material behavior of each bar $e$. Specifically, given a set $E_e$ in the phase space of member $e$, $\mu_e(E_e)$ is the probability that a fair test of member $e$ return a state $z_e \in E_e$. In accordance with this representation, in calculations we generate data sets member-wise from a zero-mean normal distribution that is no longer capped and whose standard deviation is held constant. Fig.~\ref{NonUniform_Anneal}a illustrates the data sets thus generated when the base model is as shown in Fig.~\ref{Truss_Model}b.

Since the probability measure $\mu_e$ is generated by {\sl adding} zero-mean normal random displacements to the base model in phase space, and since the constraint set $C$ is {\sl linear}, the conditional probability $\mu \LL C$ is itself centered on the base model. Hence, its mean value $\bar{z}$ necessarily coincides with the classical solution. This property is illustrated in Fig.~\ref{NonUniform_Anneal}b, which shows a convergence plot of error {\sl vs}.~data set size, with error defined as the distance between the max-ent Data Driven solution and the classical solution. For every data set size, the plot depicts histograms of error compiled from 30 randomly generated data set samples. As may be seen from the figure, the mean value of the histograms converges to zero with data set size, which is indicative of convergence in mean of the sampled max-ent Data Driven solutions. The rate of convergence of the mean error is computed to be of the order of $0.19$. Interestingly, this rate of convergence is considerably smaller than the linear convergence rate achieved for the capped normal noise distributions considered in the preceding section and also consistent with previous results \cite{RN481}. The slower rate of convergence may be attributable to the wider spread of the data about its mean, though the precise trade-off between convergence and uncertainty remains to be elucidated rigorously.

\subsection{General Performance Characteristics}

\begin{figure}
\begin{subfigure}[]{0.49\textwidth}
    \centering
    \includegraphics[width=\textwidth]{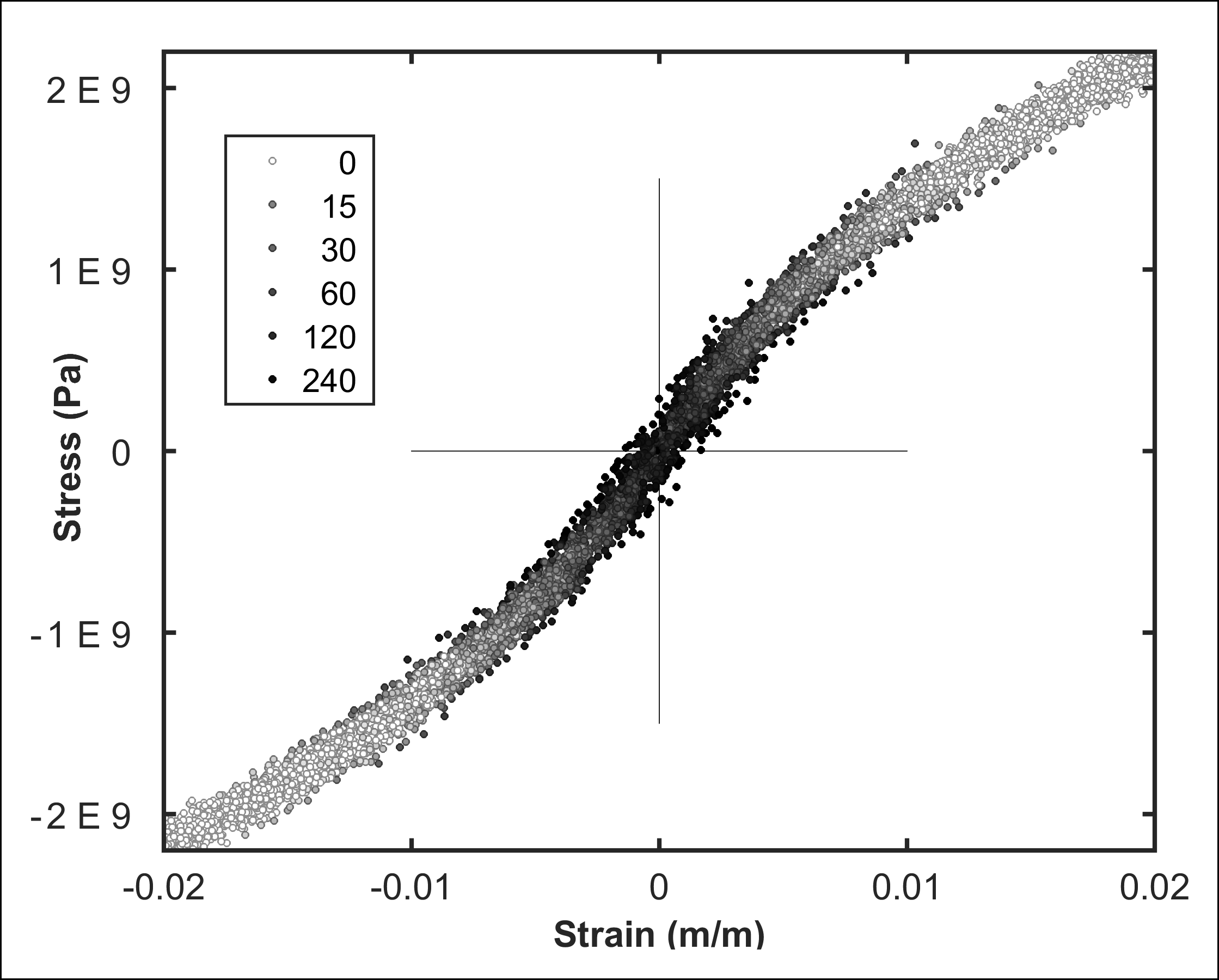}
    \caption{}
\end{subfigure}
\begin{subfigure}[]{0.49\textwidth}
    \centering
    \includegraphics[width=\textwidth]{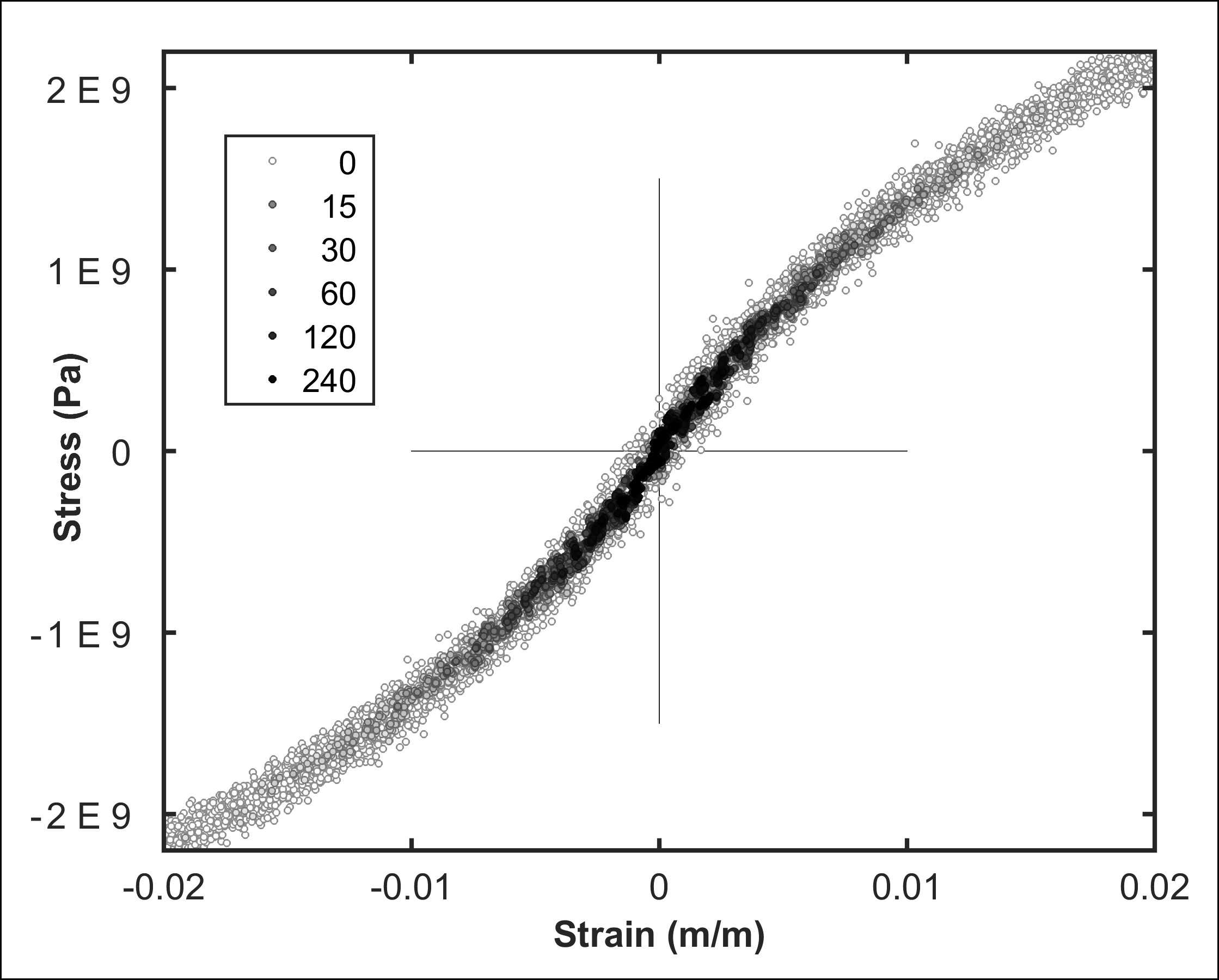}
    \caption{}
\end{subfigure}
    \caption{Data set shaded by selection frequency for a) the distance minimizing and b) entropy maximizing selection schemes.}
    \label{SelectionSet}
\end{figure}

\begin{figure}
\begin{subfigure}[]{0.49\textwidth}
    \centering
    \includegraphics[width=\textwidth]{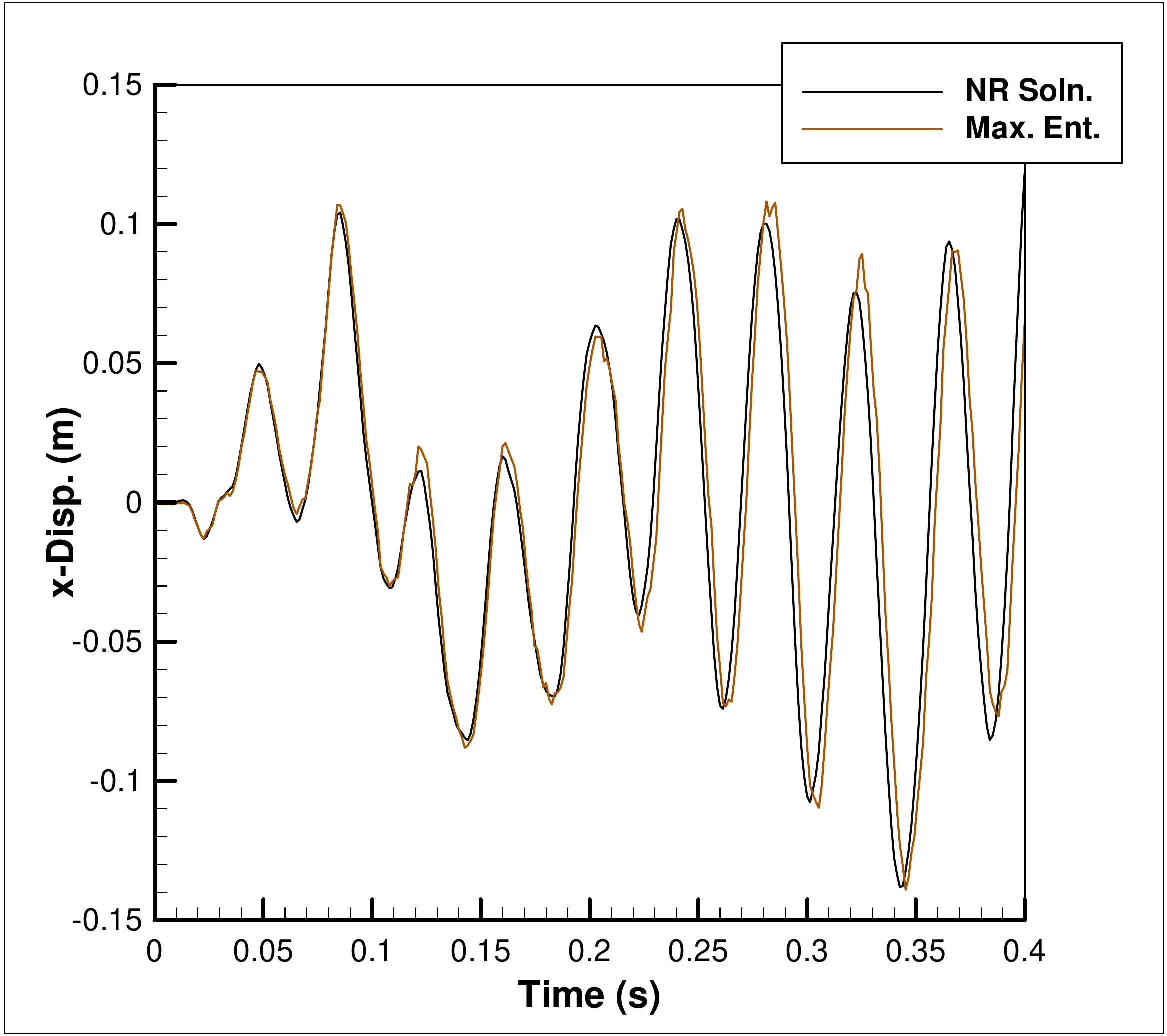}
    \caption{}
\end{subfigure}
\begin{subfigure}[]{0.49\textwidth}
    \centering
    \includegraphics[width=\textwidth]{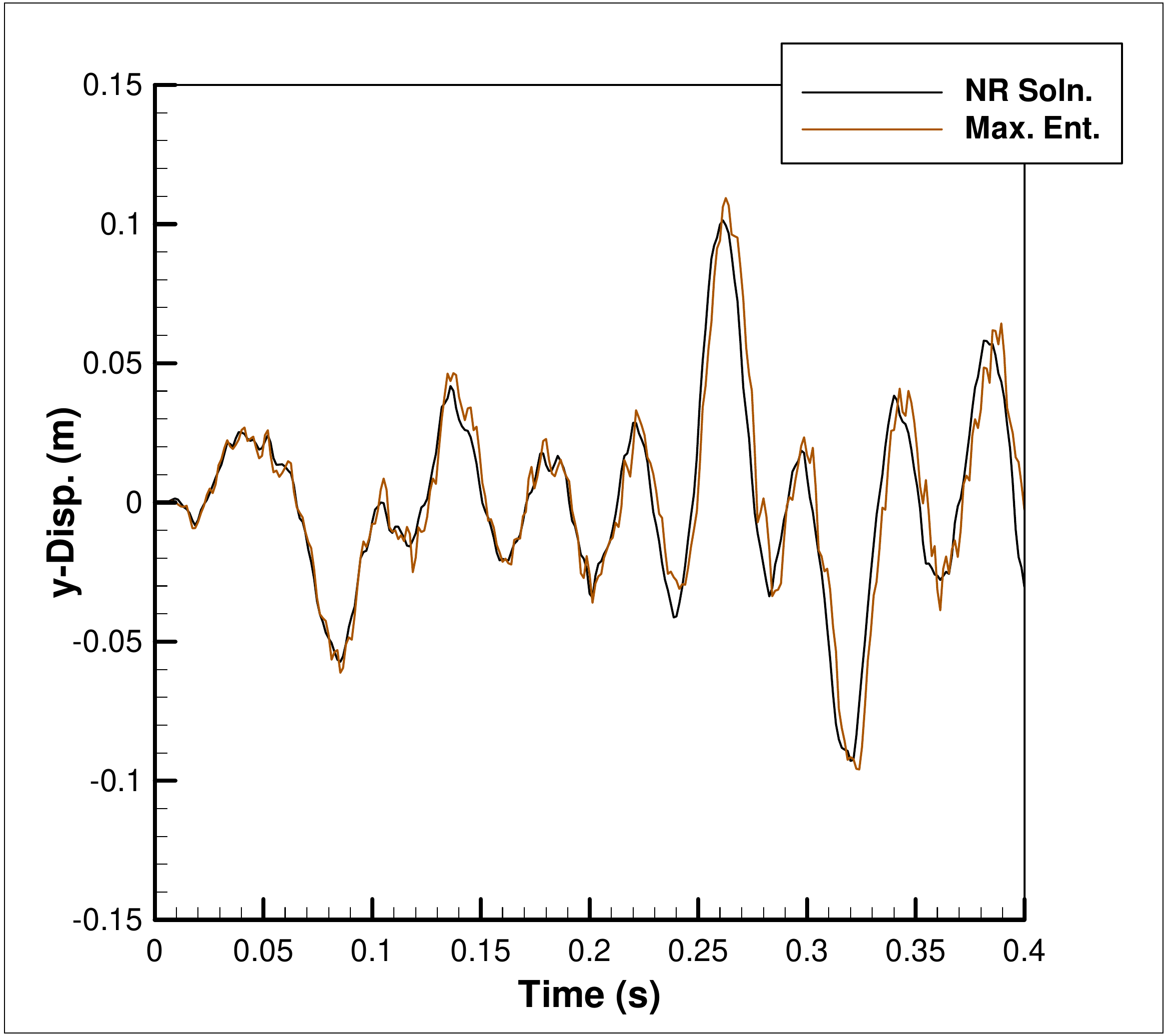}
    \caption{}
\end{subfigure}
\caption{Max-ent displacement solutions for geometry and boundary conditions seen in in Figure~\ref{Truss_Model}a solved using the data set shown in Figure~\ref{SelectionSet}  for a) the x-direction displacement and b) the y-displacement.}
    \label{DisplacementShow}
\end{figure}

Even in the presence of noiseless data sets, the max-ent solutions are uniformly seen to be improvements on distance-minimizing solutions. These significant improvements arise out of the propensity for distance-minimizing schemes to become trapped in local minima semi-adjacent to true minimizers. In turn these local data selection error accumulate with successive time integration. The benefits of entropy maximizing solutions become especially apparent in Figure \ref{SelectionSet} where the 10,000 point data set is shaded based on the number of times the various data elements were referenced in the 300 step time solution. The distance-minimizing scheme shown in Figure \ref{SelectionSet}a demonstrates how the algorithm not only allows for the selection of outliers, but how in some cases it {\sl favors} the selection of outliers. The only region where outliers are not favored is near the point $z=(0,0)$ where the elements are initialized. And while both methods were shown to converge, Figure \ref{SelectionSet} illustrates how the clustering argument made the max-ent solver robust to noisy data inputs, while the distance minimizing methods require the data converge to a graph in the phase space. Figure \ref{DisplacementShow} shows how under the clustering argument the time history of displacement maintains a remarkable fidelity to the reference displacement history for the output node.

\section{Summary and discussion}
\label{1iaFRl}

We have applied methods of Data Driven Computing paradigm, including both distance-minimizing and max-ent schemes for Data Driven Computing, to a new set of time dependent problems. We then presented selected numerical tests that establish the good convergence properties for the implemented solvers. Both distance-minimizing and max-ent solutions were shown to converge as the sequence of sets converges to an underlying model. Max-ent solutions were additionally shown to be robust to outliers and converge as the sequence of data sets converged to a fixed distribution. Beyond the specific context of improved convergence rates and conditions, max-ent solvers were also shown to have much more efficacy for step driven transient solutions because of accumulated improvements of solutions over the time domain.

An essential aspect of the Data Driven paradigm is that the space of fundamental, or model-independent, data where material data take values is determined unambiguously by the compatibility and conservation laws. This reliance on fundamental, or model-independent, material data is an essential difference with the existing Data Repositories, e.~g., \cite{NoMaD, MP, KIM, NIST}, which archive parametric data that are specific to prespecified material models. Fundamental data is {\sl fungible}, i.~e., data that is raised for one purpose can be used for another. Fundamental data is also {\sl blendable}, i.~e., material data from different sources can be blended together into a single material data set. In particular, fundamental data repositories can be publicably editable, which opens up a new and potentially far-reaching way of pooling and distributing material data.


This paper has focused on re-implementing a particular set \cite{RN48,RN481} of annealing schedules as a means to demonstrating a new class of transient Data Driven solvers. As in previous implementations, we have made no effort to speed up the implementation of the respective schedules. Thus, there remain a number of previously suggested improvements which remain unaddressed in this work e.~g. summarizing data sets, efficient range searches and radial cutoffs for summation. Time integration as affected through time stepping yields additional improvements, not implemented here, which would dramatically improve numerical performance. Specifically, previous time steps could be used to inform initialization values for an annealing process, analogous to similar strategies seen in non-linear time integration methodologies. Such initializations would allow schedules used for annealing to initiate from a $\beta^{(0)}$ larger than one which guarantees convexity over the whole data set.

\section*{Acknowledgements}
 The support of Caltech's Center of Excellence on High-Rate Deformation Physics of Heterogeneous Materials, AFOSR Award FA9550-12-1-0091, is gratefully acknowledged.

\bibliographystyle{plain}
\bibliography{biblio}

\end{document}